\documentclass{ar2e}

\pdfoutput=1
\usepackage{hyperref}
\usepackage{graphicx}
\usepackage[usenames]{color}

\def\f#1{Fig.~\ref{#1}}
\def\ff#1{Figs.~\ref{#1}}

\def\s#1{Section~\ref{#1}}

\def\c#1{~\cite{#1}}
\def\cc#1{Ref.\c{#1}}

\def\aa{(a)}

\def\cc{(c)}
\def\dd{(d)}
\def\ee{(e)}
\def\ff{(f)}

\def\as{(a) }
\def\bs{(b) }
\def\cs{(c) }
\def\ds{(d) }
\def\es{(e) }
\def\fs{(f) }

\def\beq{\begin{equation}}
\def\eeq{\end{equation}}
\def\bea{\begin{eqnarray}}
\def\eea{\end{eqnarray}}

\def\kt{k_{\rm B}T}

\begin{document}

\title{The Statistical Mechanics of Dynamic Pathways to Self-assembly}

%

%
\markboth{Whitelam and Jack}{Statistical Mechanics of Self-assembly}
\author{
  Stephen Whitelam
  \affiliation{Molecular Foundry, Lawrence Berkeley National Laboratory, 1 Cyclotron Road, Berkeley, CA 94720, USA\\  \texttt{swhitelam@lbl.gov}}
  Robert L. Jack
  \affiliation{Department of Physics, University of Bath, Bath BA2 7AY, UK \\ \texttt{r.jack@bath.ac.uk}}
}

\begin{abstract}
We describe some of the important physical characteristics of the `pathways', i.e. dynamical processes, by which molecular, nanoscale and micron-scale self-assembly occurs. We highlight the existence of features of self-assembly pathways that are common to a wide range of physical systems, even though those systems may be different in respect of their microscopic details. We summarize some existing theoretical descriptions of self-assembly pathways, and highlight areas -- notably, the description of self-assembly pathways that occur `far' from equilibrium -- that are likely to become increasingly important.
\end{abstract}

\maketitle
\eject

\section{Introduction}
 \subsection{Self-assembly}

\begin{figure*}[!h]
\includegraphics[width=1.0\linewidth]{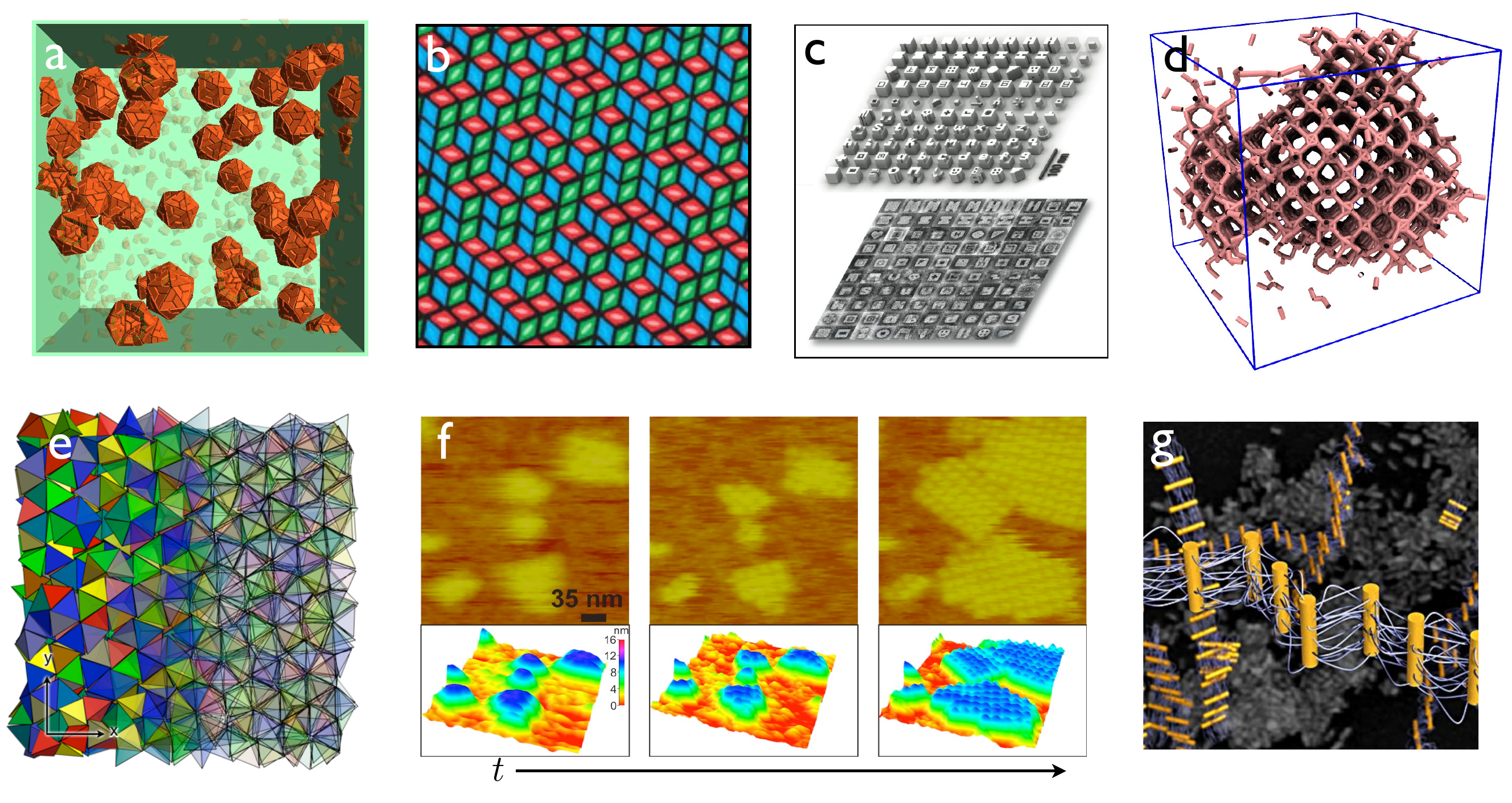}
\caption{\label{fig1} Examples of self-assembly. (a)~Model virus capsids in the process of assembling in simulation\c{rapaport2012molecular}. (b) Surface-adsorbed hydrogen-bonded network of small molecules that is statistically equivalent to a random rhombus tiling of the plane\c{blunt2008random}. (c) Multicomponent DNA `brick' structures of a range of well-defined shapes\c{ke2012three}. (d) Hierarchical rotator crystal self-assembled in simulation from clusters of spheres that bear chemically-specific interactions\c{grunwald2013patterns}. (e)  Quasicrystalline packing of tetrahedra in simulation\c{haji2009disordered}. (f)  AFM time series showing an amorphous-to-crystal self-assembly pathway of a surface-layer protein on a lipid bilayer~\cite{chung2010self}. (g) Illustration of kinetically-trapped linear structures self-assembled in experiment from DNA-linked nanorods\c{vial2013linear}.}
\end{figure*}

The term `self-assembly' describes dynamical processes in which components of a system organize themselves, without external direction, into ordered patterns or structures. The range of scales on which self-assembly happens is enormous: we might say that atoms are self-assembled from protons, neutrons and electrons, and that galaxies are self-assembled from their component stars. Here we focus on assembly undergone by components that range in size from a few angstroms (for example, atoms and molecules) to a few microns (for example, colloids). Assembly of such components is important both in the natural world and, increasingly, the laboratory~\cite{glotzer2007anisotropy,barth2005engineering,whitesides2002self,blake1999inorganic}. Structures assembled in the laboratory include three-dimensional crystals~\cite{leunissen2005ionic,zhang2005self,nykypanchuk2008dna}, two-dimensional lattices~\cite{blunt2008random,chen2011directed}, closed polyhedral shells~\cite{zlotnick2005theoretical,hagan2006dynamic}, and other tailored nanoscale shapes~\cite{rothemund2006folding,andersen2009self}. In this review, we use statistical mechanics to describe some of the pathways by which these kinds of structure are formed. Figure~\ref{fig1} displays several examples of self-assembly, taken from computer simulations and experiments.

One motivation for studying self-assembly is its potential for making new and useful materials. Since many biological systems are formed by self-assembly, we might imagine using similar processes to build new functional materials. Although we are far from this ideal, rapid progress is being made, driven by several recent advances in the synthesis of  self-assembling components.    On the molecular scale, design and selection of molecular shapes has enabled the assembly of novel structures~\cite{nam2010free,gibaud2012reconfigurable,stannard2012broken}, and DNA-mediated interactions have been used to assemble a range of complex structures~\cite{winfree1998design,park2008dna,valignat2005reversible,ke2012three}, some of which can perform basic functions~\cite{andersen2009self}.  On the colloidal scale, two key avenues for progress have been `patchy' particles, with anisotropic interactions due to chemical patterning\c{pawar2010fabrication,kraft2009colloidal,jiang2010janus}, and particles with controllable geometrical shapes\c{rossi2011cubic,sacanna2013shaping}. Imaging techniques such as atomic-force microscopy\c{yau2001direct} and in situ electron-beam methods\c{zheng2009observation,li2012direction} permit atomic-resolution imaging of some assembled structures\c{turchanin2011conversion,zhang2013recent}, and can achieve in some cases time resolution in the conditions under which self-assembly occurs\c{nielsen2014investigating,chung2010self}.

These experimental advances raise important questions for theory and modeling. Self-assembly illustrates the complex behavior accessible to simple components interacting by short-ranged forces, particularly as regards the emergence of order from disorder. Phase-ordering processes have been studied for more than a century\c{gibbs1878equilibrium,becker1935kinetic,stranski1933rate,ostwald1897studies}, and the resulting paradigm of nucleation and growth is central to our understanding of self-assembly.  However, now that assembly can be analyzed quantitatively and even visualized directly in experiments, we often find that classical theories are not sufficient to explain observed behavior\c{li2012direction,savage2009experimental}. New theoretical insights are required in order to understand and control self-assembly.

\subsection{Scope of this review}

Our aim in this review is to discuss the pathways, i.e. dynamical processes, by which self-assembly takes place. Many aspects of these pathways are generic, applying to a range of systems that may appear different in respect of their atomic or molecular details. We shall highlight some of the important `conserved' features that appear in a wide range of examples of self-assembly. We shall also stress the assumptions underpinning different theoretical descriptions of self-assembly, and the extent to which these assumptions are valid in practical settings.

 To limit the scope of our discussion, we shall restrict our focus to {\em undriven} systems, to which no energy is supplied by external means, and {\em inactive} components, whose motion is driven only by thermal fluctuations, such as those they receive from the solvent. Such components may fluctuate and undergo changes of conformation, but their motion in isolation is purely diffusive and they do not consume energy.  Pattern formation in driven systems or by active components is often called `self-organization'\c{nicolis1977self}.  Further, we shall consider only components small enough to undergo Brownian motion, which sets a size range from atoms to roughly microns. Thus, the self-assembly on which we focus is typified by an experiment in which a set of inactive components -- such as molecules, proteins, or colloids -- are dispersed in solvent, poured into a beaker, and then left alone. Given the laws of statistical mechanics, such systems can be expected to evolve toward configurations ever lower in free energy, but with no guarantee that they will achieve the thermodynamically stable state on experimental timescales.  How do we describe the fate of such systems?

To address this question, \s{physical} describes some of the key physical characteristics shared by self-assembling system.  We summarize insights obtained from several studies in terms of simple diagrams and a statistical mechanical `toy model' of self-assembly. \s{methods} contains a brief discussion of the methods used in computational modeling of self-assembly. In \s{pathway} we present some of the most important dynamic pathways seen in self-assembling systems, and \s{describe} outlines the theoretical ideas put forward to describe these pathways. Finally, \s{outlook} summarizes our outlook on the field.

\section{The physical character of self-assembly}
\label{physical}

\subsection{Thermodynamic and dynamic factors in self-assembly}
\label{physical1}

Self-assembly is a nonequilibrium process in which a system evolves from an initial disordered state toward a stable state that is usually ordered in some way. For the systems that we consider, the driving force for this process is thermodynamic in nature -- assembled structures are lower in free energy than their unassembled components -- but this does not guarantee that what self-assembles will be the equilibrium structure, or indeed a structure with special thermodynamic status.  It has been known for over a century that some components self-assemble first as a thermodynamically metastable structure, a `local minimum' of free energy distinct from the `global minimum' characteristic of the stable structure\c{ostwald1897studies,wolde1999homogeneous,cardew1985kinetics,desgranges2007controlling}. It has been known for almost as long that sometimes the structure formed first does not correspond {\em even} to a local minimum of free energy\c{stauffer1976kinetic,kremer1978multi}, although it may look ordered\c{kim2008probing}, and it need not relax to a thermodynamically-preferred structure (i.e. a minimum of free energy, local or global) an any timescale of relevance to a laboratory experiment or computer simulation.

In the following, it will be useful to distinguish two kinds of behavior: {\em near-equilibrium} assembly, in which thermodynamic factors play a dominant role, and {\em far-from-equilibrium} assembly, for which dynamic effects are vital in determining the outcome. A typical self-assembling system can display both kinds of behavior, depending on the conditions under which it is observed. Experiments and computer simulations reveal that, in a majority of examples, successful self-assembly of a stable ordered structure occurs only when system parameters are tightly controlled, so balancing two factors: a thermodynamic impetus for components to form ordered structures, and conditions that allow components moving randomly to arrange themselves into these ordered structures\c{whitesides2002beyond,hagan2006dynamic,wilber2007reversible,nguyen2007deciphering,rapaport2008role}. These factors tend to oppose each other: conditions that are optimal from a thermodynamic viewpoint are often unsuitable for dynamic reasons, and vice versa.

\begin{figure}[!h]
\includegraphics[width=\linewidth]{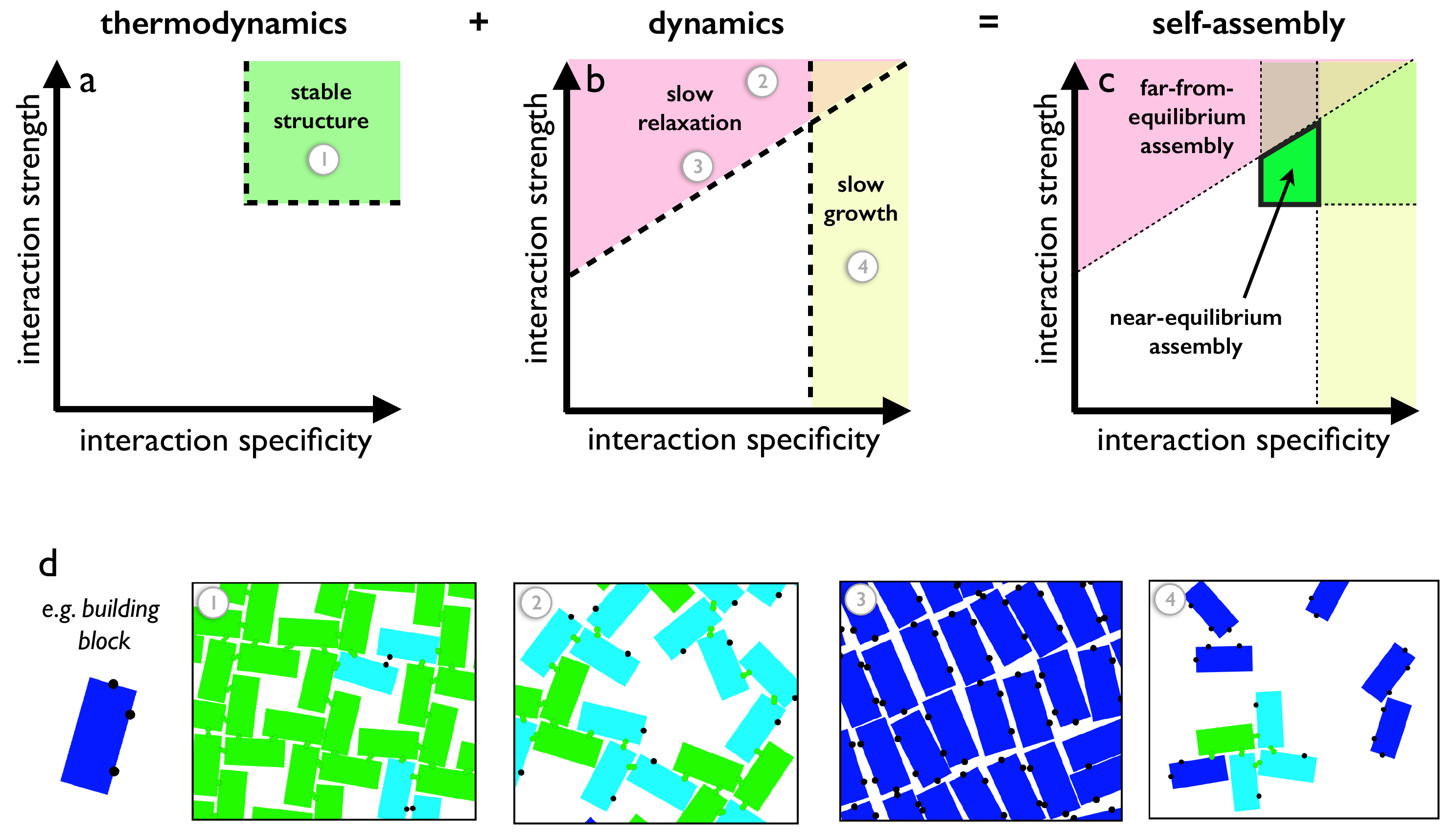} 
\caption{\label{fig2} Schematic illustration of the conflict between the requirements for the thermodynamic stability and kinetic accessibility of a desired structure, intended to summarize the collective work of many authors (see main text). Interactions between components are characterized by their strength and specificity. (a)~Schematic phase diagram, as dictated by thermodynamics. (b)~Illustration of parameter regimes in which dynamic effects dominate assembly. (c)~These competing factors result in near-equilibrium assembly occurring only for a narrow range of interaction parameters. (d)~Illustration of the structures that might be formed by an example component, at the parameter sets labeled 1--4 in (a,b). 1) Interactions with the `right' amount of strength and specificity lead to assembly of the stable ordered structure; 2) overly-strong interactions lead to kinetically-trapped structures; 3) insufficiently-specific interactions lead to alternative assembled structures; and 4) overly-specific interactions lead to no assembly. Figure adapted from Refs.~\cite{whitelam2009role,whitelam2010control}.}
\end{figure}

As an illustration of this balance, imagine that our goal, as sketched in \f{fig2}, is to design a set of components that will self-assemble in solution into a desired, thermodynamically stable structure. To ensure stability of the desired structure the interactions between components must be strong enough that an assembled structure is lower in free energy than its unassembled components. Interactions should also be `specific' in some way, so that the desired structure is lower in free energy than other possible assembled structures. Such specificity can be achieved through directional binding, as in the case of anisotropic building blocks\c{smit1993computer,damasceno2012predictive} or `patchy' nanoparticles\c{Glotzer2004patchy,pawar2010fabrication}, or through selective binding, as in the case of protein-protein interactions\c{persson2010molecular,fusco2014characterizing} or chemical complementarity\c{ke2012three,nykypanchuk2008dna,park2008dna}.

However, these requirements of strength and specificity tend to inhibit the microscopic dynamics required for successful assembly. If interactions between components are too specific, typical encounters will not result in binding, and assembly will not happen on accessible timescales.  Interactions should therefore have some characteristics that are not completely specific (indeed, many nanoscale components, such as proteins, possess nonspecific attractive forces\c{prasad2004dissection}). But nonspecific interactions lead to `mistakes': the random collision of components will not always result in geometries commensurate with the desired assembled structure. Therefore, inter-component bonds must be weak enough that `incorrect' bonds can be disrupted by thermal fluctuations. If so, components can dissociate and bind anew, in effect sampling their local environment in order to select the most favorable modes of binding. This property of microscopic reversibility is a crucial method of error-correction in self-assembly. If inter-component binding is too strong then this mechanism is suppressed, and the result is a kinetically-trapped structure~\cite{whitesides2002beyond,hagan2006dynamic,wilber2007reversible}.

This general tension between thermodynamic and dynamic factors means that assembly of a thermodynamically stable structure typically happens only in a small subset of the available parameter space (see \f{fig2}).  This is near-equilibrium assembly in the sense defined above, in which the consequences of a microscopic dynamical process can be understood in essentially thermodynamic terms. Such behavior can be seen in e.g. one-component systems near a phase boundary between ordered and disordered phases, where inter-component bonds are only moderately strong\c{wolde1999homogeneous,shen1996bcc,wolde1997epc,lutsko2006ted}.  However, one-component systems also undergo far-from-equilibrium assembly when inter-component bonds are strong, resulting in kinetically-trapped disordered structures\c{lu2008gelation}.

Theoretical work on self-assembly usually aims to determine conditions under which near-equilibrium assembly can happen\c{johnston2010modelling}, or aims to design thermodynamically stable assemblies\c{sciortino2007self,rechtsman2005optimized,rechtsman2006designed}. Understanding far-from-equilibrium assembly, though currently under-developed by comparison, may ultimately prove to be a more generally applicable strategy for building structures with desirable properties\c{rabani2003drying}, especially for systems with fundamentally slow dynamical features, such as  components with many interaction degrees of freedom\c{vial2013linear}, or collections of multiple component types\c{kim2008probing,peters2009competing}. The far-from-equilibrium regime of undriven, inactive systems will perhaps be understood using ideas similar to those used to address self-assembly driven by external fields, or of active components\c{tailleur2008statistical,ramaswamy2010mechanics}. 

\subsection{Metastable and kinetically-trapped states can persist throughout experimental observation times}
\label{physical2}

We have emphasized that both thermodynamic and dynamic factors are important in self-assembly.  The laws of thermodynamics state that an isolated system, given sufficient time, will arrive at its global free-energy minimum: it follows that the dynamic considerations of the previous section must be discussed in conjunction with the time elapsed in a self-assembly experiment or computer simulation.  To see this, consider \f{fig3}(a), which shows a  toy model~\cite{grant2011analyzing} designed to illustrate the interplay of entropic, energetic and dynamic factors inherent to self-assembly.

We consider a large number of particles, each of which can inhabit any of three microscopic states, corresponding to distinct microscopic environments.  The first state corresponds to `unbound' particles, which are free in solution. The second state corresponds to a set of $M$ `misbound' environments, in which particles make bonds that are not consistent with the final assembled structure. The third state corresponds to `optimally bound' particles, whose local binding is consistent with the thermodynamically stable assembled structure.  The energies of misbound and optimally-bound particles are $\epsilon_{\rm mis}$ and $\epsilon_{\rm opt}$ respectively: both are likely to depend on the `interaction strength' of \f{fig2}. The number of misbound states, $M$, is likely to depend on `interaction specificity'.  Particles begin in the unbound state, and rates for subsequent binding and unbinding depend on particle concentration and activation energy barriers, respectively, as shown in \f{fig3}. The equilibrium `yield' of this process, meaning the fraction of particles found in the optimally-bound environment at infinite time, is largest when the energy reward for binding is as large as possible. Equilibrium yield is shown as a black line in \f{fig3}(b) (the red arrow shows how changing the bond strength within this toy model might be related to the general scenario shown in \f{fig2}).

Dynamically, though, large binding energies act to frustrate equilibration. If bonds are strong ($\epsilon_{\rm opt},\epsilon_{\rm mis}\gg \kt$), then the system evolves rapidly to a configuration in which the fraction of optimally-bound particles is only $1/(1+M)$ (because misbound states are more accessible dynamically than is the optimal state). The basic timescale on which a system reaches its equilibrium yield is $\exp\left(\epsilon_{\rm mis}/\kt\right)$, which is large when $\epsilon_{\rm mis}$ is large. For real systems in which many incorrect bonds are made, the collective breaking of those bonds may be prohibitively slow. Given a fixed observation time, the observed yield is a non-monotonic function of interaction strength, because the increased thermodynamic driving force to populate the optimally-bound environment is counteracted by the slowness of escaping from misbound environments. Hence, yield also depends on observation time: see the solid and dashed lines on panel (b). This basic phenomenology is seen in many examples of self-assembly\c{hagan2006dynamic,wilber2007reversible,rapaport2008role}.

\begin{figure*}[!h]
\includegraphics[width=\linewidth]{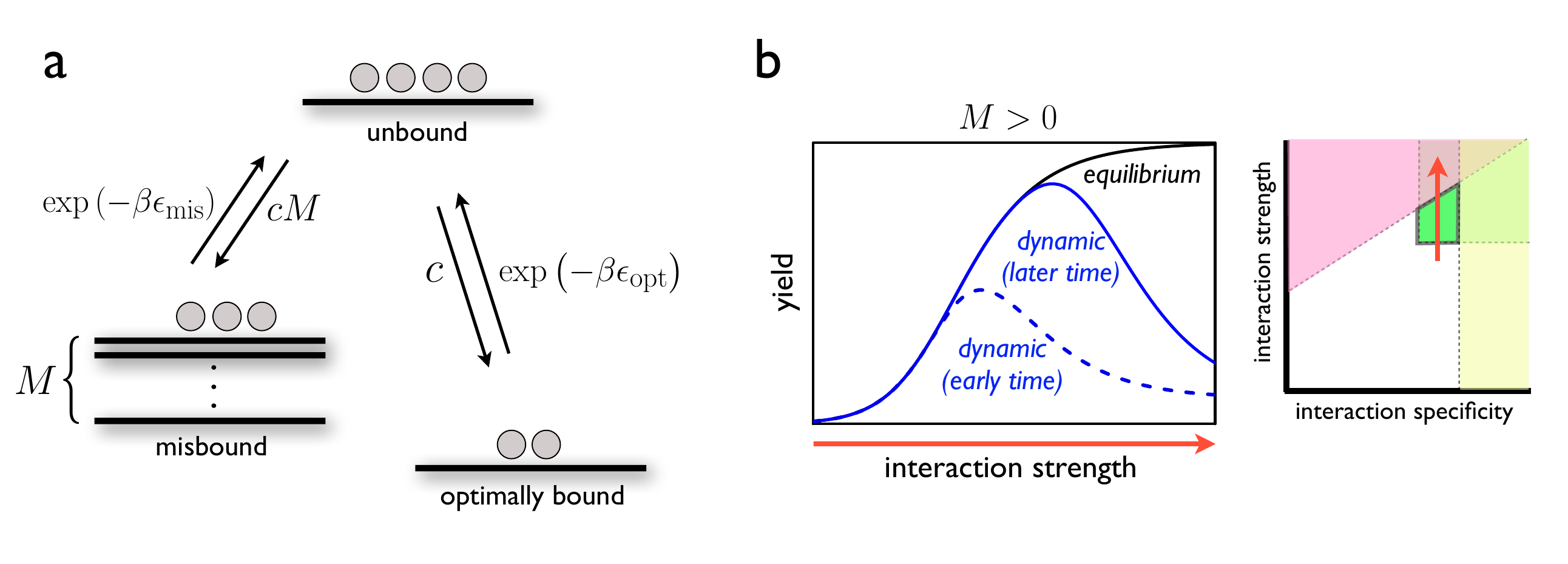}
\caption{\label{fig3} Toy statistical mechanical model of self-assembly, demonstrating the dependence of outcome on observation time, as well as the the basic requirements for kinetic trapping. (a) Particles (circles) transfer between the three microscopic environments with the rates shown. (b) When there exists the possibility of misbinding $(M>0)$, the dynamic yield depends on observation time, and at fixed time is a non-monotonic function of interaction strength. Figure adapted from Ref.\c{grant2011analyzing}.}
\end{figure*}

\section{Numerical methods for the study of self-assembly}
\label{methods}

Computer simulations play an important role in studies of self-assembly.  Components with precisely-designed interactions can be made straightforwardly on the computer, even if the experimental synthesis of the corresponding component is difficult. Following the motion of many particles in detail is simple within a computer; obtaining similar data from experimental systems is much more challenging. Here we provide a brief survey of some computer methods for the study of self-assembly.

In accordance with the preceding discussion, we separate methods into a `thermodynamic' category, which provide information about equilibrium behavior and free energies, and a `dynamic' category, designed to model the assembly process.  The standard tools of molecular simulation~\cite{frenkelsmit} are widely used: in the thermodynamic category, one typically uses Monte Carlo or molecular dynamics simulations, which may be combined with thermodynamic integration to arrive at phase diagrams\c{vega2008determination}. Advanced Monte Carlo algorithms have been developed in order to facilitate sampling of complex systems in equilibrium~\cite{bruce2003computational,liu2004rejection}. In systems that display many possible ordered structures, special algorithms can be useful for identifying candidate structures for phases~\cite{filion2009efficient}, or to identify component interactions able to stabilize chosen structures\c{rechtsman2005optimized,rechtsman2006designed}.

To obtain dynamical information about self-assembly, molecular simulation is again widely used. However, to reproduce accurately the Brownian motion of nanoscale particles, it is necessary to include an explicit representation of the solvent in which particles are dispersed.  This is computationally expensive, and so effort has been devoted to the development of methods in which solvent is treated implicitly, either through the inclusion of random forces, as in Brownian dynamics~\cite{frenkelsmit}, or through collective-move Monte Carlo methods~\cite{babu2008influence,bhattacharyay2008self,whitelam2011approximating}. In some cases, explicit and implicit solvent models show similar behavior, at least qualitatively\c{rapaport2008role,hagan2006dynamic}. In other cases, an accurate representation of solvent effects may be important, e.g. if hydrophobic effects drive assembly~\cite{wolde2002drying}, or if hydrodynamic effects are important~\cite{spaeth2011comparison,roehm2014hydro,radu2014solvent}. Given a microscopic dynamical model, rare-event-sampling methods such as forward-flux sampling~\cite{allen2006simulating} and transition-path sampling\c{wolde2002drying} are valuable if assembly involves a rare but short-lived event, such as nucleation~\cite{valeriani2005rate,sanz2007evidence}.

Whichever method is chosen, computational models of self-assembly are approximate, coarse-grained representations of experiments, typically involving approximate `effective interactions' between particles and a highly-simplifed model of solvent. Such effective interactions may be derived systematically, in a multi-scale approach~\cite{mladek2012quant,fusco2014characterizing}; fit to experimental data~\cite{abascal2005general,ouldridge2010dna}; or simply chosen in order to reproduce qualitative features of experiments~\cite{hagan2006dynamic,whitelam2010control}. Although fully quantitative agreement with experiment is difficult to achieve, especially when considering dynamical quantities~\cite{auer2001prediction}, simple models can provide useful qualitative insight into self-assembly~\cite{zhang2005self,wilber2007reversible,hagan2006dynamic,PhysRevX.4.011044,desgranges2007polymorph}.

\setcounter{section}{3}

\section{Dynamic pathways to self-assembly}
\label{pathway}

\subsection{Pathways can be near equilibrium or far from it}

\s{physical} illustrates why, in general, thermodynamic and dynamic factors must be considered in order to understand self-assembly. The message derived from several studies is that even when the interactions between components are chosen so that a particular ordered structure is stable, assembly of this structure may not be observed in experiments or computer simulations. Thus, full understanding of self-assembly requires consideration of both the desired structure and its assembly pathway~\cite{miller2010exploiting,glotzer2012pathways,klotsa2013controlling}.

In this section we discuss a range of self-assembly pathways. By `pathway' we mean a description of structures that self-assemble as time progresses. Several important pathways are shown in \f{master}, whose graphics, inspired by Fig. 2 of Ref.~\cite{wolde1997epc}, are cartoons of the objects -- monomers, stable phases, kinetically-trapped structures, etc. -- that appear during self-assembly.  A given set of components may self-assemble via more than one pathway, depending on model parameters or environmental conditions\c{de2003principles,zhang2011novel}. Shown at right in \f{master} are schematic free-energy profiles, with the free energy difference $\Delta G$ presented as a function of `progress' along each pathway. In many cases it is useful to think of the progress coordinate as the size of an assembled structure. The barrier (maximum) in free energy seen in each panel is a generic feature of phase change, familiar from classical nucleation theory~\cite{oxtoby1992homog,sear2007nucleation}: growing a cluster of a stable bulk phase results in a free energy reward that scales as the volume of the cluster, but incurs a free energy cost that scales with its surface area. Only for clusters larger than a `critical' size does reward outweigh cost, and will a cluster grow spontaneously. 

\begin{figure*}[!h]
\centering
\includegraphics[width=\linewidth]{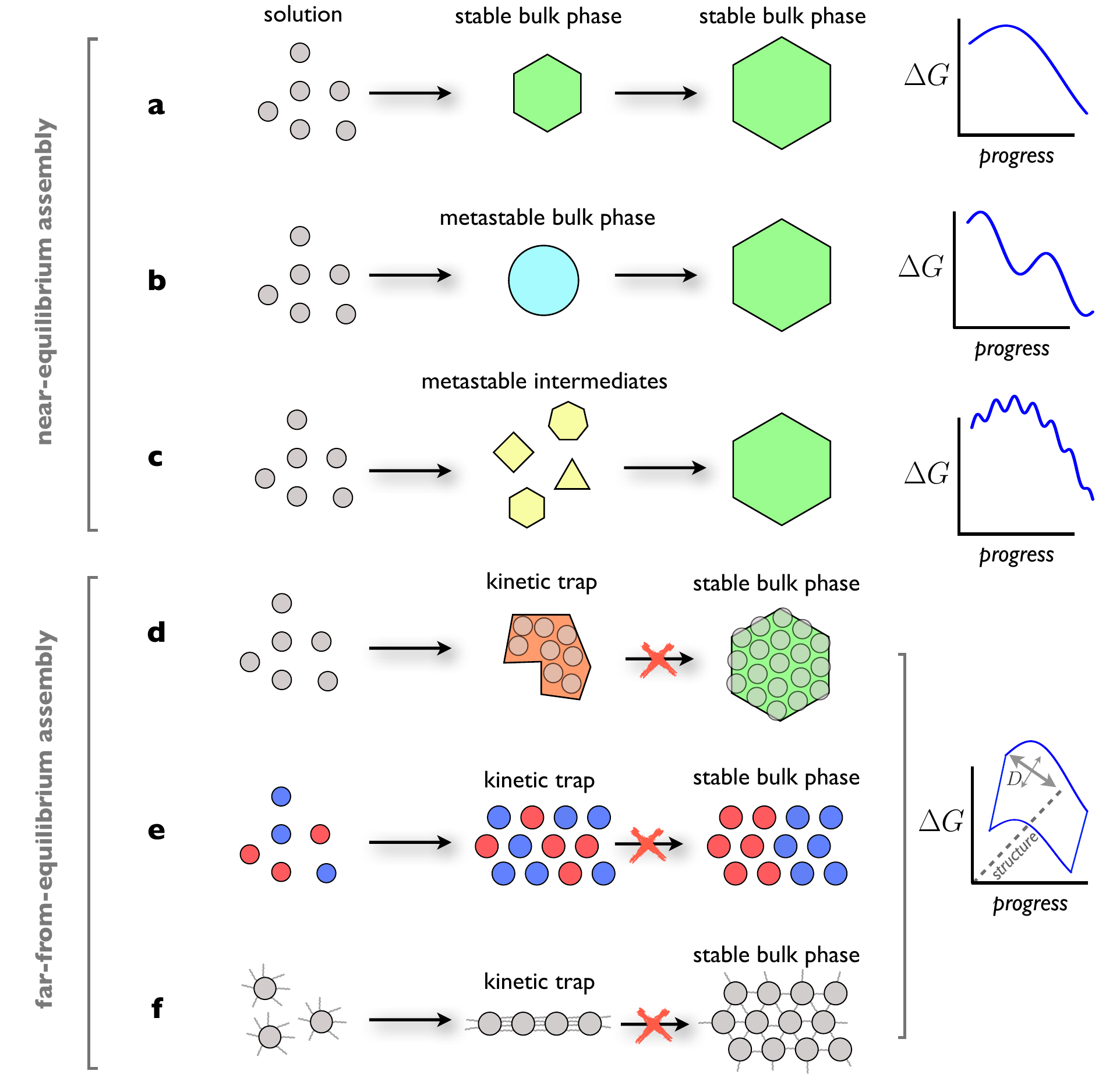}
\caption{\label{master} Examples of self-assembly pathways that are `near' equilibrium ($\aa$ to $\cc$), in the sense that they result in structures with special thermodynamic status, and `far from' equilibrium ($\dd$ to $\ff$), in the sense that they can be described only with reference to the microscopic dynamics undergone by building blocks (indicated by the ``microscopic'' detail in the pictures). These pathways are described in Sections \ref{pathway} and \ref{describe}. Figure graphics are inspired by Ref.\c{wolde1997epc}.}
\end{figure*}

We divide pathways into two general categories: near-equilibrium pathways, shown in panels \as to \cc~of \f{master}, and far-from-equilibrium pathways, shown in panels \ds to \ff. Near-equilibrium pathways can be understood in terms of evolution on a thermodynamic free-energy surface, with the dynamics of the system serving only to convey it along favored paths on that surface. By contrast, far-from-equilibrium pathways can be understood only by considering dynamic effects explicitly.  These effects result from the microscopic motion of assembling particles, hence the more `microscopic' nature of the cartoons in panels \ds to \ff. Far-from-equilibrum pathways typically involve a competition between several slow timescales, leading to motion on the underlying free-energy surface that is strongly biased by dynamic effects\c{peters2009competing}. This bias is indicated in Fig.~\ref{master} by an anisotropic diffusion tensor $D$ on the free-energy surface. The theoretical underpinnings of the `near-or-far' distinction are discussed in \s{near-far}.  

\subsection{A survey of some important self-assembly pathways}

In the remainder of this section we shall survey the pathways of \f{master}. Panel \as illustrates a near-equilibrium pathway in which components self-assemble into a thermodynamically stable structure, via the nucleation and growth of clusters with the same properties as the stable structure. This is the scenario anticipated by classical nucleation theory\c{gibbs1878equilibrium,oxtoby1992homog,sear2007nucleation}, in which the crossing of a single free-energy barrier is the rate-limiting step to formation of the stable phase. Such `classical' or `single-step' pathways have been inferred in a variety of physical systems, e.g.\c{zhang2011novel}, and have been seen with molecular-scale resolution in a few cases\c{yau2001direct}. In simulation, classical pathways are seen in the Ising lattice gas in bulk\c{ryu2010validity} and at surfaces\c{winter2009monte}, in patchy colloid models\c{chen2011directed,romano2012pattern,romano2011two}, and in atomic crystal self-assembly~\cite{valeriani2005rate}. Methods used to describe such pathways are discussed in \s{sec:path-a}.

\f{master}\bs describes a near-equilibrium pathway in which the transformation between unassembled components and the stable assembled structure is `indirect', or `multi-step', occurring via clusters whose microscopic structures are representative of thermodynamically metastable bulk phases.  An important example is the `two-step' liquid-to-crystal pathway observed during crystallization of spheres with isotropic short-range interactions~\cite{wolde1997epc}, and during the crystallization of proteins\c{vekilov2005two,chung2010self}. Multiple  transformations between metastable solid polymorphs are sometimes seen\c{chung2008multiphase}. 
Model systems with anisotropic (``patchy'') interactions can exhibit such behavior\c{wilber2007reversible,liu2009self,hedges2011limit}, as can simple lattice models~\cite{duff2009nucleation}. Methods used to describe such pathways are discussed in \s{sec:path-b}: in this pathway, the schematic `progress' coordinate in \f{master} typically includes information about the microscopic structure of the assembling cluster.

\f{master}\cs describes a near-equilibrium pathway whose intermediate structures are selected by the free-energy surface but are not directly related to {\em bulk} thermodynamic phases. Structured free-energy surfaces of this nature can result generically from faceting exhibited by finite-size clusters when component interactions are strong\c{shneidman2003lowest}, or from directional interactions that result in preferred geometries for small clusters~\cite{hagan2006dynamic,villar2009self,grunwald2013patterns}. Assembly of this nature is often hierarchical, with thermodynamically-preferred clusters serving as the building blocks for larger structures. Examples of this pathway type can be found in models and experiments of virus capsid self-assembly~\cite{hagan2014modelling,williamson2011templated}, and in the assembly of extended structures in computer simulations\c{villar2009self,grunwald2013patterns,haxton2013hierarchical}. Methods used to describe such pathways are discussed in \s{sec:path-c}. 

\f{master}\ds describes a far-from-equilibrium pathway in which components form structures that have no special thermodynamic status. The most familiar examples are the malformed, kinetically-trapped structures that result when component interactions are strong, and binding errors fail to anneal on the timescale of observation. Although kinetically-trapped structures are usually regarded as undesirable, some have interesting or useful properties: consider gels~\cite{zaccarelli2007colloidal,lu2008gelation}, fractal diffusion-limited aggregates~\cite{meakin1983formation}, or nonperiodic networks\c{PhysRevX.4.011044}. Methods used to describe such pathways are discussed in \s{sec:path-d}. 

\f{master}\es illustrates a far-from-equilibrium pathway in a system of more than one component type. Here, the physical structure may be ordered (and may be similar to the equilibrium one), but the arrangement of the particle types within that structure is not consistent with equilibrium. Multicomponent alloys display dynamically-dominated self-assembly pathways\c{clouet2006complex}; two-component colloidal crystals in experiment\c{kim2008probing} or on the computer\c{sanz2007evidence,peters2009competing} can be self-assembled with nonequilibrium component-type arrangements. The slow mobility of components within a solid structure prevents the equilibration of these arrangements on the timescale of observation. 

\f{master}\fs describes a far-from-equilibrium pathway in which kinetic trapping occurs because particles' internal degrees of freedom relax too slowly. An important example in this category is DNA-linked particles\c{wu2013kinetics} when linkers sample their configuration space more slowly than structures grow\c{vial2013linear}. Conformation change of proteins or synthetic particles can also allow particles' internal dynamics to effect dynamic control of self-assembly pathways\c{nguyen2010reconfigurable,whitelam2009impact}. We comment in \s{sec:path-ef} on the methods used to describe this pathway type, and the type sketched in panel \ee.

In addition to these well--characterized pathways, there exist several pathways seen in experiments that have yet to be classified. For instance, the thermodynamic status of clusters seen during self-assembly of mineral phases\c{gebauer2008stable,wallace2013microscopic} or some proteins\c{gliko2007metastable,sleutel2014role} is not yet clear. The fact that recently-developed experimental techniques allow molecular-scale, temporal resolution of these pathways should be seen as an exciting challenge for theory.

\section{Statistical mechanical descriptions of self-assembly pathways}
\label{describe}

\subsection{Near-equilibrium assumptions}
\label{near-far}

As discussed in Sec.~\ref{physical1}, we classify `near-equilibrium' pathways as those in which thermodynamic factors govern the outcome of an assembly process. From a theoretical perspective, the idea of a near-equilibrium pathway relies on the existence of a small number of collective co-ordinates that are sufficient to describe the assembly process.  (Similar considerations arise in the definition of reaction coordinates~\cite{hanggi1990reaction} for chemical reactions or rare events.) A natural collective co-ordinate for self-assembly is often the size $n$ of an assembling cluster\c{ten1996numerical,sear2007nucleation}, while other variables, denoted here by $m$, might describe the composition or shape of that cluster.  Let $(n_t,m_t)$ be the values of the collective coordinates in the assembling system at time $t$. In a `near-equilibrium' self-assembly process, the assembling system should have the same properties as an equilibrium system in which $(n,m)$ are constrained to be equal to $(n_t,m_t)$. That is, if two microscopic configurations of the assembling system have the same values of $n$ and $m$, and differ in energy by an amount $\Delta E$, then the ratio of the probabilities with which these configurations are seen should be ${\rm e}^{-\Delta E/(\kt)}$.  This is an example of a \emph{quasiequilibrium} condition: the only deviations of the system from equilibrium can be accounted for through the reaction co-ordinates $n,m$, with other degrees of freedom remaining equilibrated.  On time scales short enough that $(n_t,m_t)$ do not change significantly, the system then behaves as if it were at equilibrium.   Quasiequilibrium conditions hold for the `reversible' processes of classical thermodynamics\c{van1973fundamentals}.

Many theoretical descriptions of self-assembly employ a quasiequilibrium assumption, choosing a few reaction coordinates on which to focus. Such strong assumptions greatly simplify the resulting analysis. In systems with a single type of component, quasiequilibrium can be expected to hold if bond formation and bond breaking both occur rapidly on the timescale of cluster growth. The importance of reversibility and quasiequilibrium ideas in rationalizing the outcomes of self-assembly has been noted in many studies~\cite{whitesides2002beyond,hagan2006dynamic,wilber2007reversible,rapaport2008role}. The link between quasiequilibrium conditions and successful assembly has also been tested explictly in simple models~\cite{hagan2011mechanisms}. The conjecture of Stranksi and Totomanow~\cite{stranski1933rate}, that a system will transform most rapidly into the phase that requires crossing of the lowest free-energy barrier, can be justified by a quasiequilibrium assumption. When the quasiequilibrium assumption is not valid, explicit dynamical information is required in order to describe the assembly pathway.

\subsection{Classical nucleation theory}
\label{sec:path-a}

Panel \as of \f{master} represents a near-equilibrium pathway that can be described by classical nucleation theory (CNT)\c{gibbs1878equilibrium,oxtoby1992homog,sear2007nucleation}.  This theory  assumes that phase change happens via the rare nucleation of clusters, and that the structures of these clusters mimic the structure of the bulk assembled phase. As described in several review articles\c{oxtoby1992homog,sear2007nucleation}, the free-energy cost for generating a cluster of size $n$, $\Delta G(n)$, is assumed in the simplest forms of CNT to be $\Delta G(n) = \gamma n^{2/3} - n \Delta \mu $, where $\Delta \mu$ is the bulk free-energy change for formation of the stable phase, and $\gamma$ is proportional to the surface tension between the starting phase and the stable phase. The resulting free-energy barrier, $\Delta G(n^\star) \propto \gamma^3/(\Delta \mu)^2$, enters the rate for nucleation per unit volume, $k_{\rm nuc} = k_0 \exp[-\Delta G(n^\star)/\kt]$, where $k_0$ is a microsopic rate.  If $\Delta G(n^\star)$ is large compared to $\kt$, this nucleation step is expected to control the rate for assembly of an ordered phase.

CNT is a valuable starting point for describing self-assembly, especially of crystals\c{valeriani2005rate,romano2012pattern}. If intermediate states on the assembly pathway have the same kind of order as the stable state, one can expect CNT to provide at least a useful qualitative picture. Quantitative prediction of assembly rates by CNT is much rarer\c{ten1996numerical,auer2001prediction,lechner2011role}, partly because small uncertainties in calculated free-energy barriers translate into large uncertainties in nucleation rate, and because critical clusters are often not the spherical droplets assumed by simple versions of CNT\c{sear2012non}. The essential features of CNT-like ``assembly''  can be reproduced by the Ising model\c{maibaum2008ising,ryu2010validity,ryu2010numerical}, but even in this controlled setting one requires additions to the simple CNT assumptions described above in order to have quantitative agreement between theory and simulation\c{ryu2010validity,ryu2010numerical,prestipino2014shape}.  

\subsection{Beyond CNT: more than one collective coordinate}
\label{sec:path-b}

Pathway \bs of Fig.~\ref{master} illustrates a scenario in which an assembly process begins with the formation of clusters whose structure is different from that of the final assembled state. The paradigmatic example of this pathway occurs in crystallization of attractive spherical particles~\cite{wolde1997epc}, which can assemble into clusters of a metastable liquid phase during assembly; the crystal then nucleates within the liquid clusters.  As a result, assembly of the crystal occurs much more quickly than would be anticipated for a `direct' CNT pathway~\cite{wolde1997epc}.  In experiment, an important example of `two-step' assembly is seen in protein crystallization\c{vekilov2005two,sear2007nucleation}. 

Many authors have generalized the simple CNT free-energy argument so as to describe include the possibility of such a pathway: instead of a free energy $\Delta G(n)$ that depends only on cluster size, one considers a free-energy surface $\Delta G(n,m)$, in which $m$ is a measure of cluster properties (crystallinity, for instance).  Instead of a single nucleation barrier at some critical cluster size $n^\star$, one should consider the saddle points on the free-energy surface that separate the unassembled and assembled free-energy minima.  This general problem falls within the framework of multidimensional reaction-rate theory\c{hanggi1990reaction,agarwal2013solute}. 

Different assembly pathways are then expected as the shape of the free-energy surface changes.  For example, in a `double-nucleation' two-step process, assembly occurs via a pathway that passes through two saddle points, the first of which might correspond to nucleation of liquid clusters, and the second being crystal nucleation within the liquid.  In other `two-step' pathways, a single nucleation process may lead to a cluster with one kind of order, followed by the appearance of a different kind of order, as the cluster grows. For example, nuclei in systems of attractive spherical particles may be largely unstructured~\cite{wolde1997epc} or have a body-centred cubic (bcc) structure~\cite{wolde1999homogeneous}, but on long times, the system forms a face-centered cubic (fcc) crystal (without any subsequent nucleation event). Similar pathways have been found for `patchy' particles and for particles with other anisotropic interactions~\cite{wilber2007reversible,hedges2011limit,fusco2013crystallization}.

From a theoretical perspective, classical density functional theory, which assumes that the `direction' of phase change is governed by the shape of the free-energy surface, provides a suitable description of two-step self-assembly pathways, within the quasiequilibrium assumption\c{lutsko2010recent,lutsko2006ted,shen1996bcc}. Ostwald's rule of stages\c{ostwald1897studies} is the assumption that multi-step assembly will happen if there exist bulk phases intermediate in free energy between the parent phase and the stable assembly. Although often upheld\c{wolde1999homogeneous}, the statement has no theoretical underpinning, and is not predictive. Simple systems such as Potts models~\cite{sanders2007competitive,duff2009nucleation} can display two-step `assembly' pathways, helping to identify molecular features that dispose systems toward multi-step assembly. In general, one expects metastable phases to appear during phase change if particles possess a certain range of interaction\c{wolde1997epc}, or possess different types of microscopic interactions that stabilize distinct condensed phases\c{liu2009self,whitelam2010control,hedges2011limit}. 

\subsection{Assembly via specific structured clusters}
\label{sec:path-c}

Panel \cs of \f{master} describes near-equilibrium assembly which occurs via structured intermediates that combine into larger assemblies. A natural way of describing this pathway type is through a set of kinetic rate equations~\cite{binder1976statistical,becker1935kinetic} for the fission and fusion of clusters of specific morphologies. The dynamical quantities in these equations are the concentrations (number densities) of the various clusters;  the equations also include rate parameters that are determined partially by kinetic considerations and partly by thermodynamic factors (for example, detailed balance relations). Such rate equations have been used extensively to study viral capsid assembly~\cite{zlotnick1999theoretical,zlotnick2005theoretical,hagan2014modelling}: in such cases one typically assumes a dominant assembly pathway, by considering a single cluster morphology for any given size. Comparison of theory and experiments has demonstrated that such pathways occur during virus capsid assembly in vitro~\cite{hagan2014modelling}, and rate equation approaches also give a good description of the assembly of of amyloid fibrils~\cite{knowles2009analytical}, which are one-dimensional protein filaments known to cause a variety of degenerative diseases.

This pathway is also relevant for some examples of hierarchical self-assembly in which components first assemble into structured clusters, which then combine to form a larger assembly.  Examples of structured clusters might be dimeric or tetrameric protein complexes\c{villar2009self}, or `micelle-like' clusters formed from amphiphiles~\cite{miller2009hierarchical}. The essential distinction between pathways (a) and (c) in \f{master} is that a theoretical description of pathway (c) must account explicitly for the presence of small clusters with specific morphologies: a simple picture of monomer addition to a growing `droplet' of the ordered phase is not sufficient.

\subsection{Far-from-equilibrium assembly in one-component systems}
\label{sec:path-d}

We now turn to far-from-equilibrium pathways, for which a small number of collective coordinates are no longer sufficient to describe assembly (see panel \ds of \f{master}). In these situations, a crucial question is how one accounts for the variation in morphology (shape) among clusters of a given size. In principle, one may generalize the rate equation approach of the previous section, but instead of considering the concentrations of clusters of different sizes, one must consider separately clusters of all possible morphologies~\cite{binder1976statistical}. Practically speaking, the enumeration of all these possibilities is an intractable task, so the rate-equation approach is of limited applicability. Instead,
we have sketched schematic free-energy surfaces in panels \dd--\fs of \f{master}, where we have indicated the role of cluster morphology by a schematic ``structural'' axis, as a proxy for the complex range of possible cluster morphologies.  If changes in the structure of growing clusters happen quickly compared to cluster growth, one may neglect the structural axis and we recover the quasiequilibrium pathways of \as to \cc.  The far-from-equilibrium regime corresponds to the opposite limit, where clusters grow quickly enough that their structures cannot relax to their (local) equilibrium state, leading to assembled states that do not minimize the system's free energy.

The most common manifestation of these far-from-equilibrium effects is sketched in the third panel of Fig.~\ref{fig2}(d-2), where strong bonds between particles prevent the formation of ordered structures. Quantitative links between the breakdown of quasiequilibrium and the degree of order in assembled states have been confirmed by computer simulations. For example, in~\cite{hagan2011mechanisms} the quasiequilibrium assumption was tested directly in the assembly of model viral capsids.  It was found that effective assembly was associated with weak deviations from quasiequilibrium, and that the kinetic trapping regime was associated with a breakdown of quasiequilibrium. Other measurements of reversibility and quasiequilbrium can be achieved by counting events in which
clusters increase or decrease in size~\cite{rapaport2008role,grant2011analyzing} or by using relations between out-of-equilibrium correlation and response functions~\cite{jack2007fluctuation,grant2012quantifying}. In practice, successful self-assembly of an ordered structure typically involves a trade-off between cluster growth that is rapid enough for extended assemblies to form, but slow enough to achieve quasiequilibrium.  Self-assembly may involve several stages (for example, nucleation and growth): it may be that particle interactions that are optimal for one stage of assembly may be less effective in other stages.  In these situations, time-dependent interactions may be useful for optimizing assembly~\cite{klotsa2013controlling}.

We also note that while the aim of self-assembly is often to create an equilibrium structure, typically via a near-equilibrium pathway, far-from-equilibrium assembly may also be useful. Single-component systems of strongly attractive particles may form gels -- disordered networks that percolate throughout the system~\cite{zaccarelli2007colloidal,lu2008gelation}, leading to rigid (or viscoelastic) macroscopic behaviour.  Gelation is an example of a far-from-equilibrium assembly process with important applications~\cite{mezzenga2005understanding}.  In contrast to near-equilibrium assembly, the structure of assembled gels depends strongly on dynamic effects~\cite{fortini2008crystal,malins2012role}, and the assembled structures also undergo aging (dynamic effects that persist on long times), which can result in large-scale structural rearrangements~\cite{teece2011ageing}. Magnetic nanoparticles can self-assemble into nonequilibrium loop structures quite unlike the patterns expected in thermal equilibrium, highlighting the richness of morphology that can be obtained far from equilibrium~\cite{ku2010self}. The effects of dynamic factors on far-from-equilibrium assembly and gelation are not understood in detail: this remains an area in which theory and modeling have the potential to yield new insights. 

\subsection{Far-from-equilibrium assembly in more complex systems}
\label{sec:path-ef}

We now turn to panel \es of \f{master}, which illustrates systems with more than one kind of component.  In contrast to pathway \dd, where the quasiequilibrium regime breaks down because of  strong bonds between particles, the presence of multiple component types can lead to kinetic traps that emerge even when bonds between particles are weak. In a solid (e.g. a crystal) containing two component types, components interchange their positions only very slowly (if at all), and so the arrangement of component types within the assembled crystal is likely to `remember' the process by which the crystal was formed. The resulting arrangement will not in general correspond to a free-energy minimum\c{kim2008probing,scarlett2011mechanistic,sanz2007evidence,peters2009competing}. Treatments of this problem have included the development of kinetic theories in which rate parameters depend on the underlying microscopic particle dynamics\c{kremer1978multi,stauffer1976kinetic,PhysRevB.27.7372,schmelzer2004nucleation,scarlett2011mechanistic}. CNT can in principle be modified to describe similar examples of far-from-equilibrium self-assembly, accounting for slow internal relaxation of a growing cluster through the rates of change of different co-ordinates for the free-energy surface $\Delta G(n,m)$ [recall \s{sec:path-b}].  However, these rates are not calculable within CNT: they must be obtained independently, for example by computer simulation.  This
approach has been used to describe the dynamics of assembly of two-component colloidal crystals\c{peters2009competing}. Moreover, structures formed away from equilibrium may have no special thermodynamic status, and well-defined `phases' may not exist. Arguments of nonequilibrium statistical physics can provide qualitative predictions for the nature of kinetically-trapped multicomponent structures within simple models\c{whitelam2014self}, but we possess, in general, limited understanding of this phenomenon.

Polydisperse colloidal systems can also be regarded as multicomponent systems, whose components differ in size. With sufficient polydispersity the stable equilibrium state may involve coexistence of two or more crystals~\cite{sollich2011poly}, each composed of particles of a particular size range. However, phase separation or fractionation typically happens so slowly that crystal coexistence cannot be achieved on the timescale of an experiment, with systems either forming a single crystal or remaining in a disordered ``glassy'' state~\cite{sear98phase}. 

Multicomponent systems can also be made to self-assemble in quasiequilibrium\c{grunwald2013patterns,ke2012three}. Given that multicomponent self-assembly can happen near and far from equilibrium, and the fact that natural functional materials are generally multicomponent ones, there appears to be an enormous parameter space within which to design interesting self-assembled multicomponent structures, stable and kinetically trapped. Such design awaits guidance from new developments. 

The kinetic trapping seen in panel \fs of \f{master}, in which nonequilibrium structures form because of slow sampling of interaction conformations, has been seen in experiments\c{lee2009reversible,vial2013linear} and simulations\c{nguyen2010reconfigurable,whitelam2009impact}, but awaits a full microscopic dynamic description. This should be seen as a challenge to the community: given the usefulness of e.g. DNA as a mediator of interactions in self-assembly\c{rothemund2006folding,winfree1998design,ke2012three,park2008dna,nykypanchuk2008dna,valignat2005reversible}, and the slowness with which DNA linkers can sample their conformational space\c{wu2013kinetics}, it is possible that the kind of kinetic trapping seen in\c{vial2013linear} could be further developed so as to allow assembly of functional nonequilibrium structures. 

\section{Outlook}
\label{outlook}

There exist generic features of self-assembly pathways that are seen in a wide range of physical systems, even though these systems may appear different with respect to their microscopic details.  A selection of these features are summarized above, as are some of the existing theoretical descriptions of self-assembly pathways.
 In general, we possess as a community an understanding of several important general principles that apply to self-assembly. Simple theories (e.g, CNT, kinetic rate equations) can capture the qualitative behavior of many examples of self-assembly, and in some cases can give us quantitative understanding of self-assembly. Simple model systems, including Ising- and Potts-like models, model colloids, and a wide range of `patchy particle' models, have been used to reproduce the complex behavior seen in real systems without accounting for all of their microscopic details.

The foundations of the community's description of self-assembly rest on well-developed near-equilibrium ideas, and there is clear need for continuing the development of theories that are fundamentally dynamic in nature\c{lutsko2012dynamical}. Far-from-equilibrium self-assembly is likely to occur in a larger regime of parameter space than is near-equilibrium assembly; it can result in functional assemblies; and it is likely to connect naturally with intrinsically nonequilibrium phenomena like driven systems and active matter\c{ramaswamy2010mechanics}. We therefore anticipate that theoretical guidance for the far-from-equilibrium regime of self-assembly will prove increasingly important, motivated by ongoing developments in component synthesis and in-situ imaging of self-assembly pathways.

\section{Acknowledgements}

We thank Paddy Royall and Nigel Wilding for comments on the manuscript. We thank the many colleagues with whom we have discussed self-assembly, particularly Phillip L. Geissler, David Chandler, and Michael F. Hagan. S.W. performed work at the Molecular Foundry at Lawrence Berkeley National Laboratory, and was supported by the Office of Science, Office of Basic Energy Sciences, of the U.S. Department of Energy under Contract No. DE-AC02-05CH11231.  R.L.J. was funded by the Engineering and Physical Sciences Research Council (EPSRC) through grant EP/I003797/1.
\eject


\end{document}